\documentclass[aps,pre,reprint,superscriptaddress,twocolumn,longbibliography,floatfix]{revtex4-2}

\usepackage{empheq}
\usepackage{graphicx}
\usepackage{dcolumn}
\usepackage{bm}

\usepackage[utf8]{inputenc}
\usepackage[T1]{fontenc}
\usepackage{mathptmx}
\usepackage{etoolbox}
\usepackage{xcolor}
\usepackage{mathpazo}
\usepackage{bm}
\usepackage[unicode=true,
 bookmarks=true,bookmarksnumbered=false,bookmarksopen=false,
 breaklinks=false,pdfborder={0 0 0},pdfborderstyle={},backref=false,colorlinks=true]
 {hyperref}
\hypersetup{pdfborderstyle={},pdfborderstyle={},pdfborderstyle={},pdfborderstyle={},pdfborderstyle={},pdfborderstyle={},linkcolor=blue,citecolor=blue}

\usepackage{natbib}
\usepackage{amssymb,amsmath}

\newcommand\beq{\begin{equation}}
\newcommand\eeq{\end{equation}}
\newcommand\beqa{\begin{eqnarray}}
\newcommand\eeqa{\end{eqnarray}}
\newcommand{\nn}{\nonumber\\}
\def\bal#1\eal{\begin{align}#1\end{align}}
\newcommand{\Tb}{T_{\text{b}}}
\newcommand{\TbI}{T_{\text{b,I}}}
\newcommand{\TbII}{T_{\text{b,II}}}
\newcommand{\Tbm}{T_{\text{b}}^-}
\newcommand{\TbmA}{T_{\text{b,A}}^-}
\newcommand{\TbmB}{T_{\text{b,B}}^-}
\newcommand{\Tbp}{T_{\text{b}}^+}
\newcommand{\Tm}{T_{\text{b}}^{-*}}
\newcommand{\Tmm}{\hat{T}_{\text{b}}^{-*}}

\newcommand{\Tp}{T_{\text{b}}^+}
\newcommand{\TA}{T_{\text{A}}}
\newcommand{\TB}{T_{\text{B}}}
\newcommand{\TbA}{T_{\text{b,A}}}
\newcommand{\TbB}{T_{\text{b,B}}}
\newcommand{\Th}{\Tb^{\text{h}}}
\newcommand{\Tc}{\Tb^{\text{c}}}
\newcommand{\Tw}{\Tb^{\text{w}}}
\newcommand{\tw}{t_{\text{w}}}

\newcommand{\ttwmax}{\tilde{t}_{\text{w}}^{\max}}
\newcommand{\ttwmin}{\tilde{t}_{\text{w}}^{\min}}
\newcommand{\tx}{t_{\times}}

\newcommand{\taub}{\sigma}

\newcommand{\wEE}{\widetilde{\mathcal{E}}}
\newcommand{\wFF}{\widetilde{\mathcal{F}}}

\newcommand{\EE}{\mathcal{E}}
\newcommand{\FF}{\mathcal{F}}

\makeatletter
\def\@email#1#2{%
 \endgroup
 \patchcmd{\titleblock@produce}
  {\frontmatter@RRAPformat}
  {\frontmatter@RRAPformat{\produce@RRAP{*#1\href{mailto:#2}{#2}}}\frontmatter@RRAPformat}
  {}{}
}%
\makeatother
\begin{document}

\title{Time-delayed Newton's law of cooling with a finite-rate thermal quench: Impact on the Mpemba and Kovacs effects}
\author{Andr\'es Santos}%
 \affiliation{Departamento de F\'isica, Universidad de Extremadura, E-06006 Badajoz, Spain and Instituto de Computaci\'on Cient\'ifica Avanzada (ICCAEx), Universidad de Extremadura, E-06006 Badajoz, Spain}
\email{andres@unex.es}

\date{\today}

\begin{abstract}
The Mpemba and Kovacs effects are two notable memory phenomena observed in nonequilibrium relaxation processes. In a recent study [Phys.~Rev.~E \textbf{109}, 044149 (2024)], these effects were analyzed within the framework of the time-delayed Newton's law of cooling under the assumption of instantaneous temperature quenches. Here,  the analysis is extended to incorporate finite-rate quenches, characterized by a nonzero quench duration $\sigma$. The results indicate that a genuine Mpemba effect is absent under finite-rate quenches if both samples experience  the same thermal environment during the quenching process. However, if $\sigma$ remains sufficiently small, the deviations in the thermal environment stay within an acceptable range, allowing the Mpemba effect to persist with a slightly enhanced magnitude. In contrast, the Kovacs effect is significantly amplified, with the transient hump in the temperature evolution becoming more pronounced as both the waiting time and $\sigma$ increase. These findings underscore the importance of incorporating finite-time effects in nonequilibrium thermal relaxation models and offer a more realistic perspective for experimental studies.
\end{abstract}

\maketitle

\section{Introduction}
\label{sec1}

The Mpemba effect originally refers to the counterintuitive observation that, under certain conditions, hot water can freeze faster than cold water \cite{MO69,K69,F71,D71,F74,G74,W77,O79,F79,K80,H81,A95,K96,CK06,J06,K09,B11,WCVN11,BT12,S15,BT15,R15,IC16,aristotle_works_1931,Bacon_Opus_Majus,Bacon1620,Descartes1637}.

Despite extensive research, a definitive explanation of the physical mechanisms responsible for the Mpemba effect remains elusive. Proposed explanations range from evaporation effects  and convective heat transfer to differences in dissolved gas content, the influence of supercooling, or the role of surface interactions in nucleation processes
\cite{K69,F71,D71,W77,F79,WOB88,A95,M96,ERS08,K09,VM10,VM12,ZHMZZZJS14,VK15,JG15,IC16,BH20,SHZMW23}.
At the same time, skepticism remains regarding whether the effect is genuinely present in water \cite{BL16,BH20,ES21}.

Beyond its historical context, Mpemba-like behavior has been reported in a wide range of systems. For a recent review, see Ref.~\cite{TBLRV25}. Particularly, there has been growing interest in its manifestation in quantum systems, as highlighted in Refs.~\cite{TBLRV25,ACM25}.

A simple yet instructive framework for studying thermal relaxation processes is Newton's law of cooling \cite{Newton1701,B12,D12},
\beq
\label{Newton}
\dot{T}(t)=-\lambda\left[T(t)-\Tb\right],
\eeq
where \(\Tb\) is the bath temperature, and \(\lambda\) is the heat transfer coefficient. The value of \(\lambda\) depends on multiple factors, including material properties, geometry, and surrounding conditions. In the case of water, for example, it typically ranges between \(10^{-3}~\text{s}^{-1}\) and \(10^{-2}~\text{s}^{-1}\), depending on the temperature difference and volume \cite{quickfield}.

The general solution to Eq.~\eqref{Newton} is \(T(t) = \Tb + (T_0 - \Tb) e^{-\lambda t}\), where \(T_0\) is the initial temperature. It is therefore evident that the Mpemba  effect cannot occur within the framework of Eq.~\eqref{Newton}.

To account for memory effects that could explain anomalous relaxation behaviors like the Mpemba effect, a natural extension is to introduce a time delay \(\tau\) in the cooling dynamics \cite{HMMBE23,S24}:
\beq
\label{0.4}
\dot{T}(t)=-\lambda\left[T(t-\tau)-\Tb(t)\right].
\eeq
This formulation allows the system's temperature evolution to depend on its past state rather than solely on its present deviation from equilibrium. The delay parameter \(\tau\) thus plays a crucial role in capturing nontrivial thermal relaxation dynamics. Note that in Eq.~\eqref{0.4}, we have accounted for the possibility of a time-dependent bath temperature, \(\Tb(t)\).

Another important memory effect in nonequilibrium thermodynamics is the Kovacs effect, initially observed in polymer glasses \cite{K64,KAHR79} and later in various complex systems \cite{MS04,PB10,PT14,TP14,RP14,KSI17,LGAR17,PP17b,LVPS19,SP21,MS22,PSPP23}. In a typical Kovacs experiment, a system is first equilibrated at a high temperature \(\Tbm\) and then quenched to a lower temperature \(\Tbp\), where it evolves for a waiting time \(\tw\). If the bath temperature is then suddenly shifted to \(\Tw=T(\tw)\), the system exhibits a transient response: instead of staying at $T(t)=\Tw$ for $t\geq\tw$, the temperature $T(t)$ forms a characteristic hump before eventually settling to equilibrium at $\Tw$. This effect highlights how a system's thermal history influences its relaxation behavior.

In a recent study \cite{S24}, I analyzed the Mpemba and Kovacs effects within the framework of Eq.~\eqref{0.4}, assuming instantaneous temperature quenches. The analysis showed that the Mpemba effect arises only within a specific range of the control parameters \(\tau\) and \(\tw\), and that the direct (cooling) and inverse (heating) Mpemba effects are formally equivalent. Moreover, a quantitative measure of the Kovacs effect was introduced, revealing an enhancement as both \(\tau\) and \(\tw\) increase.

While theoretical studies often assume idealized instantaneous quenches, real thermal processes typically occur over finite timescales \cite{AKKL16,TBLRV25}. To model finite-rate quenches, the bath temperature \(\Tb(t)\) can be represented as
\beq
\label{Tb_single}
T_\text{b}(t)=
\begin{cases}
T_\text{b}^-,& t\leq 0,\\
T_\text{b}^++\left(T_\text{b}^--T_\text{b}^+\right) e^{-t/\taub},&t\geq 0,
\end{cases}
\eeq
where \(\taub\) characterizes the duration of the quenching process. The limit \(\taub \to 0\) recovers an instantaneous quench.

The objective of this paper is to extend the analysis of Ref.~\cite{S24} by incorporating finite-rate quenches and assessing their impact on the Mpemba and Kovacs effects. In the case of the Mpemba effect, a key distinction arises: since the two systems being quenched experience, in general, different thermal environments during the process, perfect equivalence between them no longer holds. Nonetheless, if \(\taub\) is sufficiently small, deviations remain within an acceptable range, and the Mpemba effect persists with a slightly enhanced magnitude. For the Kovacs effect, the presence of a nonzero \(\taub\) significantly amplifies the observed response, further emphasizing the role of finite-rate processes in nonequilibrium relaxation phenomena.

The rest of the paper is organized as follows. Section~\ref{sec2} derives the general solution of the time-delayed Newton's law of cooling for a single finite-rate quench. This solution is governed by an exponential-like function, \(\EE_\taub(t)\), which depends on both \(\tau\) and \(\taub\) and plays a central role in the subsequent analysis. Section~\ref{sec3} extends the solution to the case of two consecutive quenches separated by a waiting time \(\tw\). The Mpemba effect under finite-rate quenches is revisited in Sec.~\ref{sec4}, with an examination of how the quench duration \(\taub\) influences thermal relaxation dynamics. Section~\ref{sec5} focuses on the Kovacs effect, highlighting the amplification of the transient temperature hump due to finite-time quenching. Finally, Sec.~\ref{sec6} presents the main conclusions and outlines potential directions for future research.

\section{Single quench}
\label{sec2}
For simplicity, henceforth we take $\lambda^{-1}=1$ as the unit of time.
Let us first assume that the bath temperature experiences a single quench, as described by Eq.~\eqref{Tb_single}. In order to solve Eq.~\eqref{0.4} under those circumstances, it is convenient to work in Laplace space, as done in Ref.~\cite{S24}.

The Laplace transform of Eq.~\eqref{Tb_single} is
\beq
\label{Tbtilde}
\widetilde{T}_\text{b}(s)=\Tbp s^{-1}+\frac{\Tbm-\Tbp}{s+\taub^{-1}}.
\eeq
Therefore, Eq.~\eqref{0.4} yields
\beq
\label{Ttilde}
\widetilde{T}(s)=\Tbp s^{-1}+\left(\Tbm-\Tbp\right)\wEE_\taub(s),
\eeq
with
\beq
\label{4}
\wEE_\taub(s)=\wEE_0(s)+\wFF_\taub(s),
\eeq
where
\begin{subequations}
\label{5}
\bal
\label{5a}
\wEE_0(s)\equiv&s^{-1}-\frac{s^{-2}}{1+s^{-1}e^{-s\tau}}\nn
=&s^{-1}-\sum_{n=0}^\infty (-s)^{-(n+2)}e^{-ns\tau},
\eal
\bal
\label{5b}
\wFF_\taub(s)\equiv &\frac{s^{-1}}{\left(1+s^{-1}e^{-s\tau}\right)\left(s+\taub^{-1}\right)}\nn
=&-\sum_{n=0}^\infty\frac{(-s)^{-(n+1)}e^{-ns\tau}}{s+\taub^{-1}}.
\eal
\end{subequations}
Note that \(\lim_{\taub\to 0} \wFF_\taub(s) = 0\), meaning that only the function \(\wEE_0(s)\) is relevant in the special case of an instantaneous quench.

In real time, Eq.~\eqref{Ttilde} becomes
\beq
\label{6}
T(t)=\begin{cases}
\Tbm,& t\leq 0,\\
\Tbp +\left(\Tbm-\Tbp\right)\EE_\taub(t),& t\geq 0,
\end{cases}
\eeq
where $\EE_\taub(t)=\mathcal{L}^{-1}\left[\wEE_\taub(s)\right]=\EE_0(t)+\FF_\taub(t)$. Here, $\EE_0(t)=\mathcal{L}^{-1}\left[\wEE_0(s)\right]$ is given by \cite{S24}
\beq
\label{14}
\EE_0(t)=1+\sum_{n=0}^{\lfloor{t/\tau}\rfloor} \frac{(n\tau-t)^{n+1}}{(n+1)!},\quad t\geq 0,
\eeq
where $\lfloor\cdot\rfloor$ denotes the floor function.
To obtain $\FF_\taub(t)=\mathcal{L}^{-1}\left[\wFF_\taub(s)\right]$, we need the property
\bal
\mathcal{L}^{-1}\left[\frac{s^{-(n+1)}}{s+a}\right]=&\frac{e^{-at}}{(-a)^{n+1}}+\lim_{s\to 0}\frac{1}{n!}\partial_s^{n} \frac{e^{st}}{s+a}\nn
=&\frac{R_n(at)}{(-a)^{n+1}},
\eal
where
\beq
R_n(x)\equiv e^{-x}-\sum_{k=0}^{n}\frac{(-x)^k}{k!}
\eeq
is the remainder of the exponential series.
Therefore,
\beq
\FF_\taub(t)=-\sum_{n=0}^{\lfloor{t/\tau}\rfloor}\taub^{n+1}R_n((t-n\tau)/\taub).
\eeq

Note that
\beq
\label{dotEE}
\dot{\EE}_\taub(t)=e^{-t/\taub}-\begin{cases}
1,&0\leq t\leq \tau,\\
\EE_\taub(t-\tau),&t\geq \tau.
\end{cases}
\eeq
Equation~\eqref{dotEE} implies that $\dot{\EE}_\taub(0)=0$ if $\taub\neq 0$, while $\dot{\EE}_0=-1$.

The expressions for $\EE_0(t)$, $\FF_\taub(t)$, and $\EE_\taub(t)$ in the absence of a time delay ($\tau = 0$) are detailed in Appendix \ref{appA}. Additionally, Appendix \ref{appB} provides the formulation of $\FF_\taub(t)$ in the quasi-instantaneous quench regime ($\taub \ll 1$).

\begin{figure}
      \includegraphics[width=0.9\columnwidth]{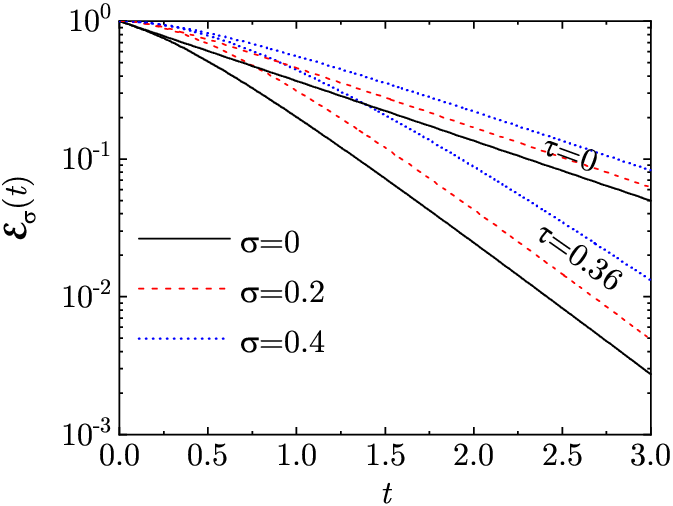}
      \caption{Plot of $\EE_\taub(t)$ for  $\taub=0$, $0.2$,  and $0.4$. The values of the delay time are $\tau=0$ (upper curves) and $\tau=0.36$ (lower curves).
  \label{fig1}}
\end{figure}

\begin{figure}
      \includegraphics[width=0.9\columnwidth]{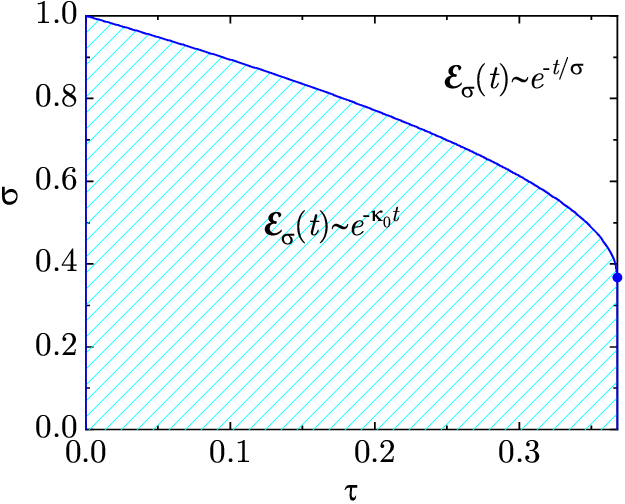}
      \caption{Plane $\taub$ vs $\tau$ showing the region where $\EE_\taub(s)$ decays asymptotically as $e^{-\kappa_0 t}$ (below the curve $\taub=\kappa_0^{-1}$) and the region where $\EE_\taub(s)$ decays asymptotically as $e^{-t/\taub}$ (above the curve). The circle denotes the end of the curve, where $\kappa_0^{-1}(\tau_{\max})=\tau_{\max}=e^{-1}\simeq 0.368$.
  \label{fig2}}
\end{figure}

As Fig.~\ref{fig1} shows, at a given delay time $\tau$, the relaxation of $\EE_\taub(t)$ becomes slower as $\taub$ increases. In the absence of time delay ($\tau=0$),  one has $\EE_\taub(t)\approx e^{-t}/(1-\taub)$ [see Eq.~\eqref{11a}], while the decay is much faster if $\tau\neq 0$.

Let us now analyze the asymptotic behavior of $\EE_0(t)$ and $\FF_\taub(t)$ for long times. From  the first line of Eq.~\eqref{5a} we see that the poles of $\wEE_0(s)$ are  the roots of the transcendental equation $s+e^{-s\tau}=0$. If $\tau<\tau_{\max}=e^{-1}\simeq 0.368$ \cite{S24}, there are two real roots, namely $s=-\kappa_0$ and $s=-\kappa_1$, where $\kappa_0=-\tau^{-1}W_0(-\tau)$, $\kappa_1=-\tau^{-1}W_{-1}(-\tau)$,  $W_0(z)$ and $W_{-1}(z)$ being the principal and lower branches of the Lambert function \cite{CGHJK96}, respectively. The remaining complex roots of $s+e^{-s\tau}=0$ have a real part more negative than $\kappa_0$ and $\kappa_1$, so they can be neglected in the asymptotic form for $\EE_0(t)$. Therefore, for asymptotically long times,
\beq
\label{AEE0}
\EE_0(t)\approx
\frac{\kappa_0^{-1} e^{-\kappa_0 t}}{1-\kappa_0\tau}-\frac{\kappa_1^{-1} e^{-\kappa_1 t}}{\kappa_1\tau-1}.
\eeq
Since $1 < \kappa_0 < e \simeq 2.718 < \tau^{-1} < \kappa_1$, the first term on the right-hand side of Eq.~\eqref{AEE0} dominates over the second in the regime $t\gg 1$. Nonetheless, the inclusion of the second term ensures that Eq.~\eqref{AEE0} excellently approximates $\EE_0(t)$ for all $t$, especially when $t > 2\tau$. The maximum deviation of the approximation given by Eq.~\eqref{AEE0} from $\EE_0(t)$ occurs at $t = 0$ and $\tau = \tau_{\max} = e^{-1}$, with a value of $\frac{8}{3}e^{-1} - 1 \simeq -0.019$.

Equation \eqref{5b} shows that the poles of $\wFF(s)$ are $s=-\taub^{-1}$ and the roots of  $s+e^{-s\tau}=0$, Therefore,
\bal
\label{AFF}
\FF_\taub(t)\approx&
\frac{\taub e^{-\kappa_0 t}}{(1-\kappa_0\taub)(1-\kappa_0\tau)}-
\frac{\taub e^{-t/\taub}}{1-\taub e^{\tau/\taub}}\nn
&-\frac{\taub e^{-\kappa_1 t}}{(1-\kappa_1\taub)(\kappa_1\tau-1)}.
\eal
The approximate form for $\EE_\taub(t)$ is then
\bal
\label{AEE}
\EE_\taub(t)\approx&
\frac{\kappa_0^{-1} e^{-\kappa_0 t}}{(1-\kappa_0\taub)(1-\kappa_0\tau)}-
\frac{\taub e^{-t/\taub}}{1-\taub e^{\tau/\taub}}\nn
&-\frac{\kappa_1^{-1} e^{-\kappa_1 t}}{(1-\kappa_1\taub)(\kappa_1\tau-1)}.
\eal
Again, this provides an excellent approximation for all $t$. Note also that, since $\lim_{\tau\to 0}\kappa_0=1$ and $\lim_{\tau\to 0}\kappa_1=\infty$, Eq.~\eqref{AEE} reduces to Eq.~\eqref{11a} in the limit $\tau\to 0$.

In the long-time limit, the leading behavior of $\EE_\taub(t)$ is determined by the first term on the right-hand side of Eq.~\eqref{AEE} if $\taub < \kappa_0^{-1}$, while the second term dominates if $\taub > \kappa_0^{-1}$.
On the line $\taub=\kappa_0^{-1}$ both competing poles coalesce into a common double pole, so that, in that case,
\beq
\label{border}
\EE_\taub(t)\approx\left[\frac{1+\kappa_0 t}{\kappa_0(1-\kappa_0 \tau)}-\frac{\kappa_0 \tau^2}{2(1-\kappa_0\tau)^2}\right]e^{-\kappa_0 t}.
\eeq
Figure \ref{fig2} illustrates the regions in the $\taub$ vs $\tau$ plane where each type of asymptotic behavior prevails. These regions are divided by the line $\taub = \kappa_0^{-1}$.
Note that  $\EE_\taub(t)$ is always larger than $e^{-t/\taub}$, even if $\taub>\kappa_0^{-1}$, implying that $\Tb(t)$ relaxes to $\Tbp$ earlier than $T(t)$, as expected. On the other hand, only if $\taub<\kappa_0^{-1}$ does $\Tb(t)$ relaxes exponentially faster than $T(t)$.

Suppose that two samples (A and B) undergo the quench described by Eq.~\eqref{Tb_single}, except that their initial bath temperatures differ, with \(\TbmA > \TbmB\), while the quench time (\(t=0\)) and final bath temperature (\(\Tp\)) remain the same for both. According to Eq.~\eqref{6}, their relative temperature difference evolves as
\beq
\label{17}
\frac{\TA(t)-\TB(t) }{\TbmA-\TbmB}=\begin{cases}
1,& t\leq 0,\\
\EE_\taub(t),& t\geq 0.
\end{cases}
\eeq
Since \(\EE_\taub(t) > 0\) (provided that \(\tau < \tau_{\text{max}}\)), Eq.~\eqref{17} demonstrates that a Mpemba effect is not possible under a simultaneous single-quench protocol, regardless of the value of \(\taub\).

\section{Double quench}
\label{sec3}

Now, instead of considering the single-quench protocol described by Eq.~\eqref{Tb_single}, let us assume a double-quench protocol, where the material is initially held at a prior temperature \(\Tbm\) for \( t < -\tw \), then quenched to an intermediate bath temperature \(\Tm\) for \(-\tw \leq t < 0\), and finally quenched again to a final bath temperature \(\Tbp\) at \( t = 0 \). The corresponding bath temperature profile is given by
\beq
\label{14b}
\Tb(t)=
\begin{cases}
\Tbm, & t\leq -\tw,\\
\Tm+\left(\Tbm-\Tm\right) e^{-(t+\tw)/\taub}, & -\tw\leq t\leq 0,\\
\Tbp+\left(\Tmm-\Tbp\right) e^{-t/\taub}, & t\geq 0,
\end{cases}
\eeq
where
\beq
\Tmm=\Tb(0)=\Tm+\left(\Tbm-\Tm\right) e^{-\tw/\taub}.
\eeq
Note that  \(\Tb(t)\) is a continuous function, which in turn guarantees continuity in \(\dot{T}(t)\), in contrast to the case of instantaneous quenches (\(\taub = 0\)).

The double quench described by Eq.~\eqref{14b} can be decomposed as the sum of two single quenches:
\beq
\Tb(t)=\TbI(t)+\TbII(t),
\eeq
where
\begin{subequations}
\beq
\TbI(t)=\begin{cases}
\Tbm-\Tm+\displaystyle{\frac{\Tbp}{2}},&t\leq -\tw,\\
\displaystyle{\frac{\Tbp}{2}}+\left(\Tbm-\Tm\right)e^{-(t+\tw)/\taub},&t\geq -\tw,
\end{cases}
\eeq
\beq
\TbII(t)=
\begin{cases}
\Tm-\displaystyle{\frac{\Tbp}{2}},&t\leq 0,\\
\displaystyle{\frac{\Tbp}{2}}+\left(\Tm-\Tbp\right)e^{-t/\taub},&t\geq 0.
\end{cases}
\eeq
\end{subequations}
From Eq.~\eqref{6}, we can write
\begin{subequations}
\beq
T_{\text{I}}(t)=\begin{cases}
\Tbm-\Tm+\displaystyle{\frac{\Tbp}{2}},& t\leq -\tw,\\
\displaystyle{\frac{\Tbp}{2}} +\left(\Tbm-\Tm\right)\EE_\taub(t+\tw),& t\geq -\tw,
\end{cases}
\eeq
\beq
T_\text{II}(t)=\begin{cases}
\Tm-\displaystyle{\frac{\Tbp}{2}},& t\leq 0,\\
\displaystyle{\frac{\Tbp}{2}} +\left(\Tm-\Tbp\right)\EE_\taub(t),& t\geq 0.
\end{cases}
\eeq
\end{subequations}
Therefore, since $T(t)=T_\text{I}(t)+T_\text{II}(t)$, the solution corresponding to the double quench given by Eq.~\eqref{14b} is
\beq
\label{T_dq}
T(t)=
\begin{cases}
\Tbm,&t\leq-\tw,\\
\Tm+\left(\Tbm-\Tm\right)\EE_\taub(t+\tw),&-\tw\leq t\leq 0,\\
\Tbp+\left(\Tbm-\Tm\right)\EE_\taub(t+\tw)\\
+\left(\Tm-\Tbp\right)\EE_\taub(t),&t\geq 0.
\end{cases}
\eeq

\begin{figure}
      \includegraphics[width=0.49\columnwidth]{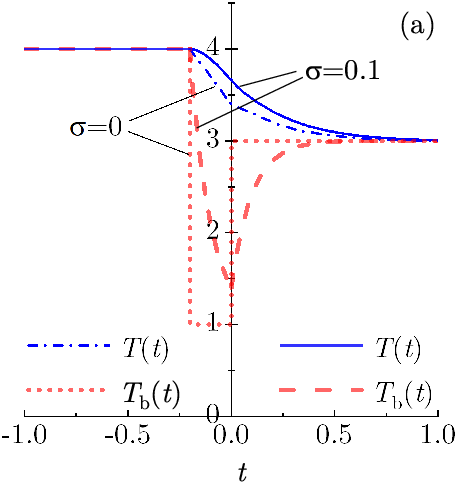}
       \includegraphics[width=0.49\columnwidth]{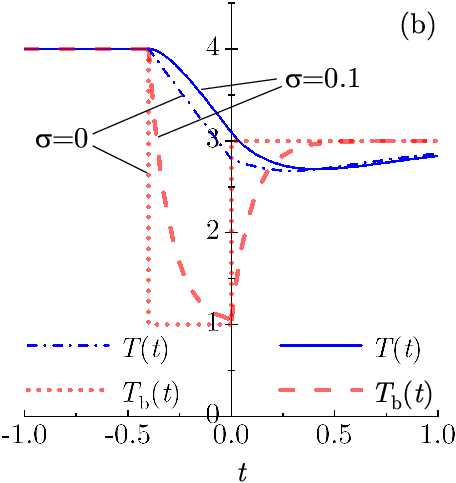}
      \caption{Sketch of a generic double finite-rate quench. Here, $\Tbm=4$ (in arbitrary units), $\Tm=1$, $\Tbp=3$, $\taub=0.1$, $\tau=0.36$, and (a) $\tw=0.2$, (b) $\tw=0.4$. The dashed lines represent the bath temperature $\Tb(t)$, while the solid lines represent the system temperature $T(t)$. For comparison, also the cases of instantaneous quenches ($\taub=0$) are included, where the dotted and dash-dotted lines represent $\Tb(t)$ and $T(t)$, respectively.
  \label{fig3}}
\end{figure}

A sketch of the double finite-rate quench is shown in Fig.~\ref{fig3}, which also includes the case of the instantaneous quench ($\taub=0$).

\begin{figure}
      \includegraphics[width=0.49\columnwidth]{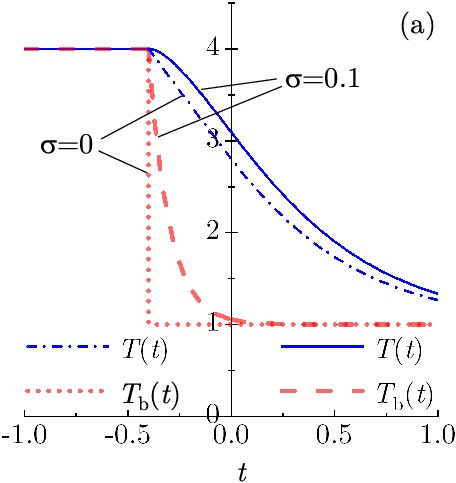}
       \includegraphics[width=0.49\columnwidth]{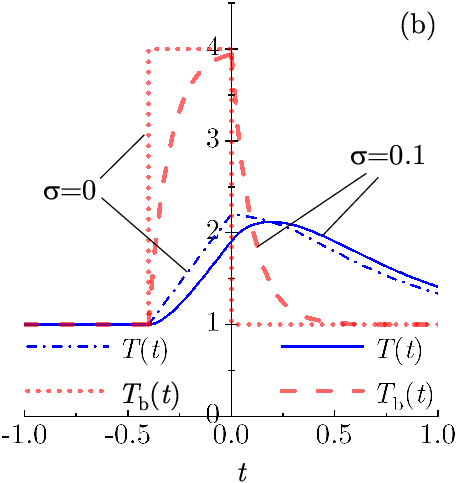}
      \caption{Illustration of the protocols (a) A and (b) B described in Table \ref{tab1}.  Here, $\Th=4$ (in arbitrary units), $\Tc=1$,  $\taub=0.1$, $\tau=0.36$, and $\tw=0.4$. The dashed lines represent the bath temperature $\Tb(t)$, while the solid lines represent the system temperature $T(t)$. For comparison, also the cases of instantaneous quenches ($\taub=0$) are included, where the dotted and dash-dotted lines represent $\Tb(t)$ and $T(t)$, respectively.
  \label{fig4}}
\end{figure}

\section{Impact of finite-rate quenches on the Mpemba effect}
\label{sec4}

\begin{table}
\caption{Protocols A and B for the Mpemba effect and protocol for the Kovacs effect. Note that in the Mpemba effect, the quenches take place at \( t = -\tw \) and \( t = 0 \), whereas in the Kovacs effect they take place at \( t = 0 \) and \( t = \tw \).\label{tab1}}
\begin{ruledtabular}
\begin{tabular}{lccc}
Protocol&$\Tbm$&$\Tm$&$\Tbp$\\
\hline
A (Mpemba)&$\Th$&$\Tc$&$\Tc$\\
B (Mpemba)&$\Tc$&$\Th$&$\Tc$\\
Kovacs &$\Th$&$\Tc$&$\Tw$\\
\end{tabular}
\end{ruledtabular}
\end{table}

To investigate the possibility of a Mpemba effect with finite-rate quenches ($\taub\neq 0$), we consider the single-quench (A) and double-quench (B) protocols summarized in Table~\ref{tab1}. Those protocols are the same as considered in Ref.~\cite{S24} with $\taub=0$. As we see, only two thermal reservoirs are considered, a hot one (at temperature $\Th$) and a cold one (at temperature $\Tc$).
The protocols A and B are illustrated in Fig.~\ref{fig4}, which displays the cases with $\taub\neq 0$ and $\taub=0$.

According to Eq.~\eqref{T_dq},
\begin{subequations}
\beq
\TA(t)=
\begin{cases}
\Th,&t\leq-\tw,\\
\Tc+\left(\Th-\Tc\right)\EE_\taub(t+\tw),&t\geq -\tw,\\
\end{cases}
\eeq
\beq
\TB(t)=
\begin{cases}
\Tc,&\hspace{-0.7cm}t\leq-\tw,\\
\Th-\left(\Th-\Tc\right)\EE_\taub(t+\tw),&\hspace{-1.3cm}-\tw\leq t\leq 0,\\
\Tc+\left(\Th-\Tc\right)\left[\EE_\taub(t)-\EE_\taub(t+\tw)\right],&t\geq 0.
\end{cases}
\eeq
\end{subequations}
Therefore, for $t\geq 0$,
\beq
\label{23}
\frac{\TA(t)-\TB(t)}{\Th-\Tc}=\Delta_\taub(t),\quad t\geq 0,
\eeq
where the difference function is
\beq
\label{Delta}
\Delta_\taub(t)=2\EE_\taub(t+\tw)-\EE_\taub(t).
\eeq
From Eq.~\eqref{AEE}, we see that the approximate, asymptotic form of $\Delta_\taub(t)$ is
\bal
\label{ASS}
\Delta_\taub(t)\approx&
\frac{\kappa_0^{-1}(2e^{-\kappa_0\tw}-1) e^{-\kappa_0 t}}{(1-\kappa_0\taub)(1-\kappa_0\tau)}-
\frac{\taub (2e^{-\tw/\taub}-1)e^{-t/\taub}}{1-\taub e^{\tau/\taub}}\nn
&-\frac{\kappa_1^{-1}(2e^{-\kappa_1\tw}-1) e^{-\kappa_1 t}}{(1-\kappa_1\taub)(\kappa_1\tau-1)}.
\eal

According to Eq.~\eqref{23}, the existence of a Mpemba effect requires (i) $\Delta_\taub(0)>0$ and (ii) $\lim_{t\to\infty}\Delta(t)<0$.

\begin{figure}
      \includegraphics[height=0.48\columnwidth]{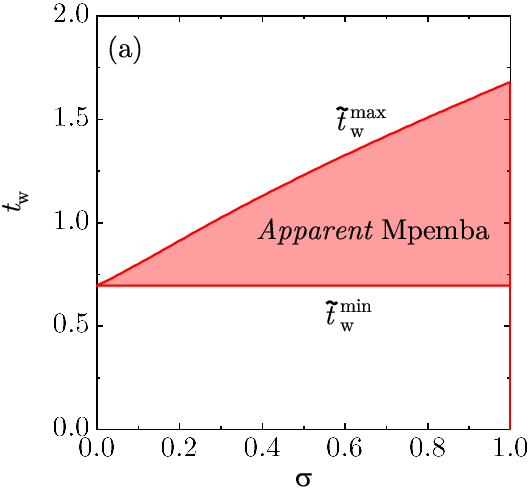} \includegraphics[height=0.48\columnwidth]{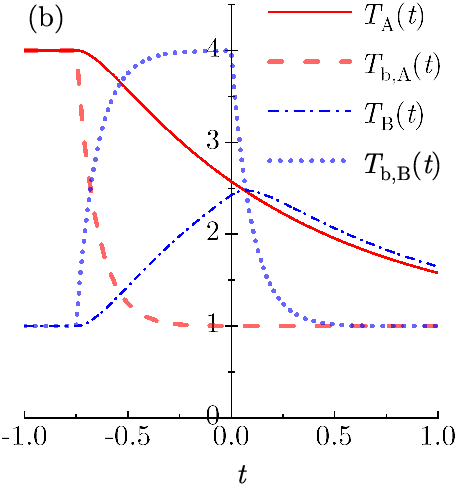}
      \caption{(a) Phase space for the apparent Mpemba effect in the absence of time delay ($\tau=0$). The lower and upper curves represent $\ttwmin$ and $\ttwmax$, respectively, as functions of the quench characteristic time $\taub$. If $\tw<\ttwmin$, then $\Delta_\taub^0(t)>0$ for $t\geq 0$;      if, on the other hand, $\tw>\ttwmax$, then $\Delta_\taub^0(t)<0$ for $t\geq 0$. Therefore, an apparent Mpemba effect occurs if and only if $\ttwmin<\tw<\ttwmax$ (shaded region). (b) Evolution of $\TA(t)$ (solid line), $\TB(t)$ (dash-dotted line), $\TbA(t)$ (dashed line), and $\TbB(t)$ (dotted line) for the undelayed case ($\tau=0$) with $\taub=0.1$, $\tw=0.75$, $\Th=4$ (in arbitrary units), and $\Tc=1$.
  \label{fig5}}
\end{figure}

\subsection{Apparent Mpemba effect with $\tau=0$ and $\taub\neq 0$}
Let us first assume Newton's cooling law without any delay time ($\tau=0$) but with finite-rate quenches ($0<\taub<1$). In that case,  from either Eq.~\eqref{11a} or Eq.~\eqref{ASS}, we have
\bal
\label{Delta_0}
\Delta_\taub^0(t)\equiv&\lim_{\tau\to 0}\Delta_\taub(t)\nn
=&
\frac{(2e^{-\tw}-1) e^{- t}-\taub (2e^{-\tw/\taub}-1)e^{-t/\taub}}{1-\taub}.
\eal
One can see that $\Delta_\taub^0(0)>0$ if the waiting time is smaller than a maximum value $\ttwmax$ given by the solution to the equation
\beq
e^{-\ttwmax}-\taub e^{-\ttwmax/\taub}=\frac{1-\taub}{2}.
\eeq
This maximum value ranges from $\ttwmax=\ln 2\simeq 0.693$ at $\taub=0$ to $\ttwmax=-W_{-1}(-e^{-1}/2)-1\simeq 1.678$ at $\taub=1$.
Furthermore, $\lim_{t\to\infty}\Delta_\taub^0(t)<0$ if the waiting time $\tw$ is larger than a minimum value $\ttwmin=\ln 2$. Within the interval $\ttwmin<\tw<\ttwmax$, $\Delta_\taub^0(t)=0$ at a crossing time
\beq
\tx=\frac{\taub}{1-\taub}\ln \frac{\taub (2e^{-\tw/\taub}-1)}{2e^{-\tw}-1}.
\eeq
Figure \ref{fig5}(a) shows the $\taub$ dependence of $\ttwmax$ and $\ttwmin$, highlighting the region where $\ttwmin < \tw < \ttwmax$.

How can a memory phenomenon, such as the Mpemba effect, arise in the \emph{undelayed} cooling equation? To resolve this apparent paradox, let us examine the case $\taub = 0.1$ and $\tw = 0.75$, which lies within the shaded region of Fig.~\ref{fig5}(a). Figure \ref{fig5}(b) shows the time evolution of the temperatures of samples A and B, along with the corresponding bath temperatures. We observe that $\TA(0) > \TB(0)$, but $\TA(t) < \TB(t)$ for $t > \tx \simeq 0.066$. This behavior arises because, unlike instantaneous quenches ($\taub = 0$), the bath temperatures influencing samples A and B for very short positive times are effectively $\Tc$ and $\Th$, respectively. Consequently, this represents an \emph{apparent} Mpemba effect, not driven by memory but by the differing environmental conditions experienced by the two samples for $t > 0$, even though the reservoir temperature is $\Tc$ in both cases.

\begin{figure}
      \includegraphics[width=0.32\columnwidth]{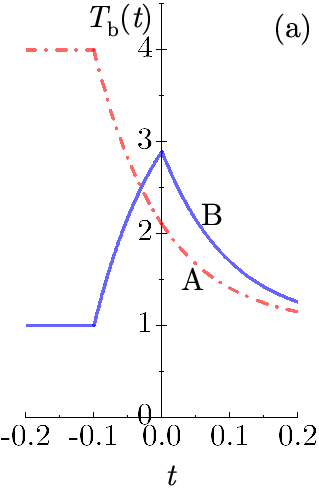}
      \includegraphics[width=0.32\columnwidth]{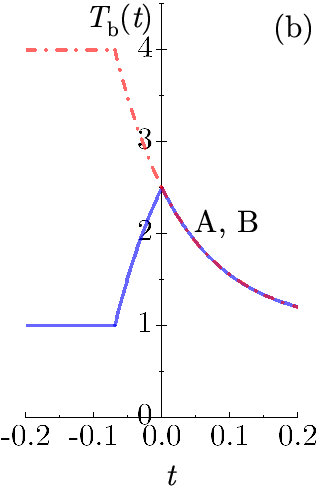}
      \includegraphics[width=0.32\columnwidth]{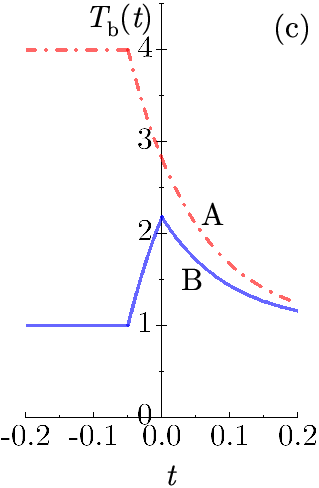}
      \caption{Bath temperatures in protocols A and B. Here, $\Th=4$ (in arbitrary units), $\Tc=1$,  $\taub=0.1$,  and (a) $\tw=0.1>\taub\ln 2$, (b)   $\tw=0.0693=\taub\ln 2$, and (c) $\tw=0.05<\taub\ln 2$.
  \label{fig6}}
\end{figure}

\subsection{Absence of a genuine Mpemba effect with $\taub\neq0$}
\label{sec4B}

From Eq.~\eqref{14b}, we see that the bath temperatures in protocols A and B for $t\geq 0$ are
\begin{subequations}
\label{TbA&TbB}
\beq
T_{\text{b,A}}(t)=\Tc+\left(\Th-\Tc\right)e^{-(t+\tw)/\taub},
\eeq
\beq
T_{\text{b,B}}(t)=\Tc+\left(\Th-\Tc\right)\left(1-e^{-\tw/\taub}\right)e^{-t/\taub}.
\eeq
\end{subequations}
Their relative difference is
\beq
\label{TbA&TbB_2}
\frac{T_{\text{b,A}}(t)-T_{\text{b,B}}(t)}{\Th-\Tc}=\left(2e^{-\tw/\taub}-1\right)e^{-t/\taub}, \quad t\geq 0.
\eeq

The concept of a genuine Mpemba effect requires that samples A and B experience the \emph{same} time-dependent bath temperature for $t > 0$. This condition is automatically satisfied when $\taub = 0$, as in this case, $T_{\text{b,A}}(t) = T_{\text{b,B}}(t) = \Tc$ for $t > 0$. However, when $\taub \neq 0$, Eq.~\eqref{TbA&TbB_2} indicates that $T_{\text{b,A}}(t) = T_{\text{b,B}}(t)$ for $t > 0$ only if the waiting time is coupled to the quench characteristic time by
\beq
\label{twa&twB}
\tw=\taub\ln 2.
\eeq
Figure~\ref{fig6} illustrates the time-dependence of the bath temperatures $T_{\text{b,A}}(t)$ and $T_{\text{b,B}}(t)$ if (a) $\tw>\taub\ln2$, (b) $\tw=\taub\ln2$, and (c) $\tw<\taub\ln2$. Only the scenario depicted in Fig.~\ref{fig6}(b) allows for a ``fair'' assessment of whether a Mpemba effect is present.

It can be verified that the difference function at $t = 0$, $\Delta_\taub(0) = 2\EE_\taub(\tw) - 1$, is always greater than $2(1 - \ln 2) \simeq 0.614$ when $\tw = \taub \ln 2$. Therefore, the presence of the Mpemba effect would require $\lim_{t \to \infty} \Delta_\taub(t) < 0$.
For $\tw = \taub \ln 2$, the second term on the right-hand side of Eq.~\eqref{ASS} identically vanishes, simplifying the asymptotic behavior of $\Delta_\taub(t)$ to
\beq
\label{ASS2}
\Delta_\taub(t)\approx
\frac{\kappa_0^{-1}(2^{1-\kappa_0\taub}-1) e^{-\kappa_0 t}}{(1-\taub\kappa_0)(1-\kappa_0\tau)}.
\eeq
Considering the mathematical property $(2^{1-x} - 1)/(1 - x) > 0$ for all $x$, we conclude that $\lim_{t \to \infty} \Delta_\taub(t) > 0$ if Eq.~\eqref{twa&twB} is satisfied. Consequently, strictly speaking, no Mpemba effect is possible when $\taub \neq 0$. In contrast, for instantaneous quenches ($\taub = 0$), a Mpemba effect occurs if $\kappa_0^{-1} \ln 2 < \tw < \tw^{\max}$, where $\tw^{\max}$ is the root of $\Delta_0(0) = 2\EE_0(\tw) - 1 = 0$ \cite{S24}.

\begin{figure}
      \includegraphics[width=0.32\columnwidth]{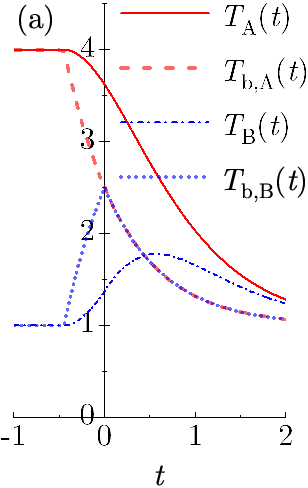}
      \includegraphics[width=0.32\columnwidth]{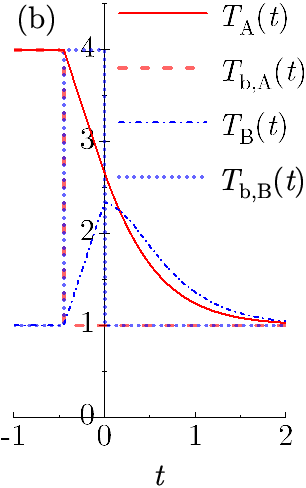}
      \includegraphics[width=0.32\columnwidth]{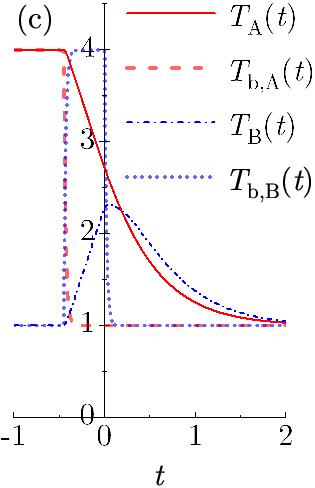}
      \caption{Bath and sample temperatures in protocols A and B. Here, $\Th=4$ (in arbitrary units), $\Tc=1$,  $\tau=0.36$,  $\tw=0.45$, and (a) $\taub=\tw/\ln 2\simeq 0.649$, (b) $\taub=0$, and (c) $\taub=0.02$.
  \label{fig7}}
\end{figure}

\begin{figure}
      \includegraphics[width=0.9\columnwidth]{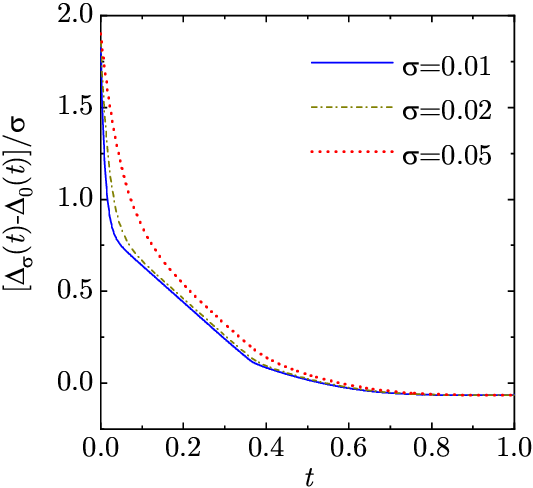}
            \caption{Plot of $\left[\Delta_\taub(t)-\Delta_0(t)\right]/\taub$ vs $t$ for  $\tau=0.36$,  $\tw=0.45$, and three values of $\taub$.
  \label{fig8}}
\end{figure}

\subsection{Admitting an imperfect Mpemba effect with small $\taub$}

As discussed in Sec.~\ref{sec4B}, the two systems A and B being quenched are not, in general, strictly subject to identical environmental conditions, as the bath temperatures differ during the quenching process for $t>0$. However, if $\taub$ is sufficiently small, the deviations in environmental conditions can be considered negligible within a certain tolerance. This approximate equivalence allows the Mpemba effect observed under instantaneous quenches to persist, albeit imperfectly.

If $\taub \ll \tw$, Eq.~\eqref{TbA&TbB_2} gives $ |T_{\text{b,A}}(t) - T_{\text{b,B}}(t)| / (\Th - \Tc) \approx e^{-t/\taub} $, which becomes smaller than a tolerance value $\epsilon$ for $ t > \taub \ln \epsilon^{-1} $. For example, if $\epsilon = 0.01$ and $\taub = 0.02$, the condition $ |T_{\text{b,A}}(t) - T_{\text{b,B}}(t)| / (\Th - \Tc) < \epsilon $ holds for $ t > 0.092 $.

Figure~\ref{fig7} illustrates the evolution of the bath and sample temperatures for $\tau = 0.36$, $\tw = 0.45$, and (a) $\taub = \tw / \ln 2 \simeq 0.649$, (b) $\taub = 0$, and (c) $\taub = 0.02$.
In Fig.~\ref{fig7}(a), $ T_{\text{b,A}}(t) = T_{\text{b,B}}(t) $ for $ t > 0 $, but no Mpemba effect is observed. Conversely, Fig.~\ref{fig7}(b) shows a clear Mpemba effect for instantaneous quenches ($\taub = 0$). In Fig.~\ref{fig7}(c), an imperfect Mpemba effect is depicted. Here, although $ T_{\text{b,A}}(t) \neq T_{\text{b,B}}(t) $ for $ t > 0 $, the relative difference $ |T_{\text{b,A}}(t) - T_{\text{b,B}}(t)| / (\Th - \Tc) $ remains below a tolerance value $\epsilon = 0.0067$ for $ t > 0.1 $.

Having established the existence of an imperfect Mpemba effect for quasi-instantaneous quenches ($\taub\ll1$), we now examine how these cases differ from instantaneous quenches ($\taub=0$). Figure~\ref{fig8} illustrates this comparison through the time evolution of $\left[\Delta_\taub(t)-\Delta_0(t)\right]/\taub$ for $\tau=0.36$, $\tw=0.45$, and three small values of $\taub$. The results reveal that, while initially $\Delta_\taub(0)>\Delta_0(0)$, at sufficiently long times the inequality reverses, yielding $\Delta_\taub(t)<\Delta_0(t)$. This suggests that when a Mpemba effect exists for $\taub=0$, its magnitude is actually enhanced (albeit slightly) for small but nonzero values of $\taub$.

\section{Impact of finite-rate quenches on the Kovacs effect}
\label{sec5}

In the Kovacs effect, a first quench from $\Th$ to $\Tc$ takes place at $t=0$. Subsequently, after a waiting time $\tw$, a second quench is performed to a reservoir at the instantaneous system temperature $\Tw \equiv T(\tw)$, as outlined in the last row of Table~\ref{tab1}.
Thus, according to Eq.~\eqref{T_dq}, the sample temperature in the  Kovacs effect is
\beq
\label{Kovacs}
T(t)=\begin{cases}
\Th,&t\leq 0,\\
\Tc+(\Th-\Tc)\EE_\taub(t),&0\leq t\leq \tw,\\
\Tw+(\Tc-\Tw)\EE_\taub(t-\tw)\\
+(\Th-\Tc)\EE_\taub(t),&t\geq \tw,
\end{cases}
\eeq
with
\beq
\label{Tw}
\Tw=\Tc+(\Th-\Tc)\EE_\taub(\tw).
\eeq

\begin{figure}
      \includegraphics[width=0.49\columnwidth]{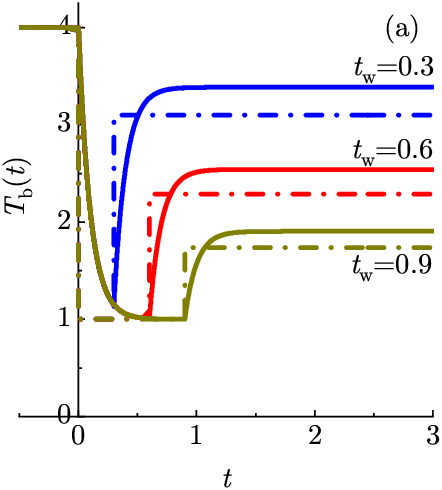}
       \includegraphics[width=0.49\columnwidth]{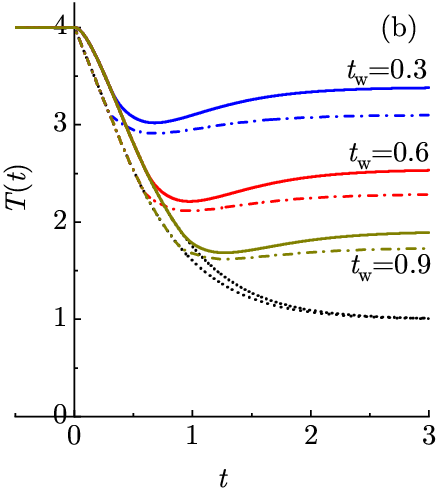}
      \caption{
      Kovacs effect for \(\Th = 4\) (arbitrary units), \(\Tc = 1\), \(\tau = 0.36\), and \(\tw = 0.3, 0.6, 0.9\), comparing instantaneous quenches (\(\taub = 0\), dash-dotted lines) and finite-rate quenches (\(\taub = 0.1\), solid lines).
(a) Bath temperature \(\Tb(t)\).
(b) Sample temperature \(T(t)\), where dotted lines represent \(T(t)\) if the system were not quenched to \(T(\tw)\) at \(t = \tw\).
  \label{fig9}}
\end{figure}

Figure \ref{fig9} compares the Kovacs effect for \(\tau = 0.36\) and three waiting times (\(\tw = 0.3, 0.6, 0.9\)) under instantaneous (\(\taub = 0\)) and finite-rate (\(\taub = 0.1\)) quenches. Three key differences are observed: (i) the temperature slope \(\dot{T}(t)\) is continuous at \(t = 0\) and \(t = \tw\) for finite-rate quenches, whereas it is discontinuous for instantaneous quenches; (ii) the sample temperature \(\Tw\) at the waiting time is higher for finite-rate quenches than for instantaneous quenches; and (iii)  the Kovacs hump is deeper in the case of finite-rate quenches.

To better quantify the latter property, we define the Kovacs hump function \( K_\taub(t) \) as
\beq
K_\taub(t )=-\frac{T(t+\tw) - \Tw}{\Tw - \Tc}, \quad t \geq 0.
\eeq
This formulation isolates the relative deviation of the sample temperature from \(\Tw\) after the quench, allowing for a clearer analysis of the hump's  magnitude.
From Eqs.~\eqref{Kovacs} and \eqref{Tw}, one finds
\beq
K_\taub(t)=\EE_\taub(t)-\frac{\EE_\taub(t+\tw)}{\EE_\taub(\tw)}.
\eeq
Note that \( K_\taub(t) \) is a non-negative function that vanishes at \( t = 0 \) and asymptotically approaches zero as \( t \to \infty \).

Using Eq.~\eqref{dotEE}, we obtain
\bal
\dot{K}_\taub(t)=&\left[1-\frac{e^{-\tw/\taub}}{\EE_\taub(\tw)}\right]e^{-t/\taub}\nn
&+\begin{cases}
\frac{1}{\EE_\taub(\tw)}-1,&0\leq t\leq \max\{0,\tau-\tw\},\\
\frac{\EE_\taub(t+\tw-\tau)}{\EE_\taub(\tw)}-1,&\max\{0,\tau-\tw\}\leq t\leq \tau,\\
-K_\taub(t-\tau),&t\geq \tau.
\end{cases}
\eal
While \(\dot{K}_0(\tau) = -K_0(0) = 0\), indicating that \(K_0(t)\) reaches its maximum at \(t = \tau\), in the case of finite-rate quenches, we have
$\dot{K}_\taub(\tau) = \left[1 - {e^{-\tw/\taub}}/{\EE_\taub(\tw)}\right] e^{-\tau/\taub} > 0$,
which implies that the maximum of \(K_\taub(t)\) occurs at a time greater than \(\tau\).

\begin{figure}
      \includegraphics[width=0.49\columnwidth]{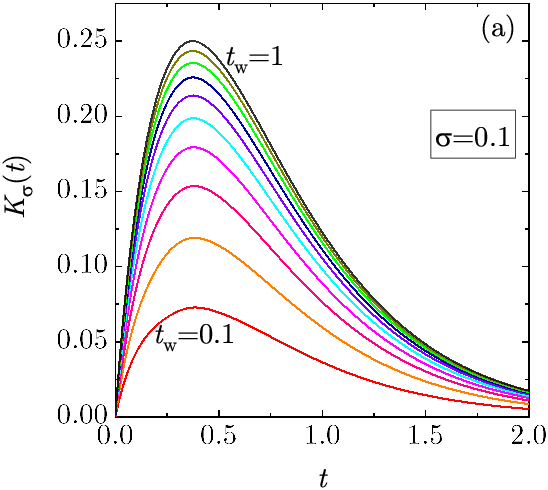}
      \includegraphics[width=0.49\columnwidth]{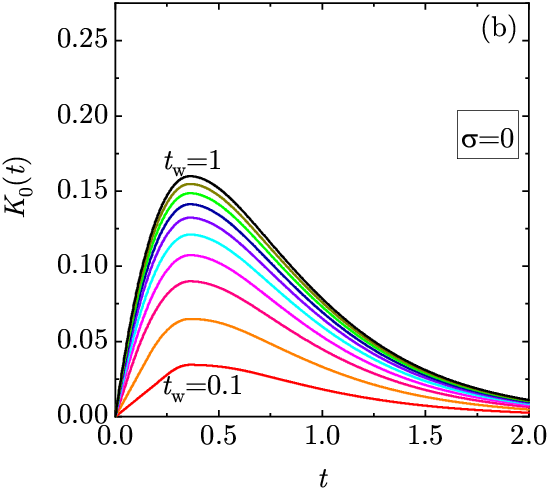}
      \caption{Plot of the Kovacs hump function \(K_\taub(t)\) for \(\tau = 0.36\) and waiting times \(\tw = 0.1, 0.2, \dots, 1\). (a) Finite-rate quenches with \(\taub = 0.1\). (b) Instantaneous quenches (\(\taub = 0\)).
      \label{fig10}}
\end{figure}

Figure~\ref{fig10} shows the Kovacs hump function \(K_\taub(t)\) for \(\tau = 0.36\) and waiting times \(\tw = 0.1, 0.2, \dots, 1\), comparing finite-rate quenches (\(\taub = 0.1\)) and instantaneous quenches (\(\taub = 0\)). In both cases, the relative magnitude of the effect increases with \(\tw\). Additionally, the presence of a nonzero characteristic quench time significantly amplifies the effect.

\section{Conclusions}
\label{sec6}

In this work, I have extended the study of the Mpemba and Kovacs effects within the framework of the time-delayed Newton's law of cooling by incorporating finite-rate thermal quenches. While previous analyses assumed instantaneous temperature changes \cite{S24}, this paper has explored the impact of a nonzero quench duration on these nonequilibrium relaxation phenomena.

The results show that, for the Mpemba effect, the presence of a finite quench time \(\taub\) introduces deviations in the environmental conditions experienced by the two initially different samples. As a consequence, the perfect equivalence that exists under instantaneous quenches no longer holds. However, if \(\taub\) remains sufficiently small, the effect persists, albeit with a slightly enhanced magnitude.

For the Kovacs effect, it has been found that a finite quench time significantly amplifies the transient hump observed in the system's temperature evolution. The effect becomes more pronounced as both the waiting time and the quench duration increase, further highlighting the role of memory in thermal relaxation.

Although the graphs illustrate only the direct (cooling) effects, the results naturally extend to inverse (heating) effects, as the functions \(\EE_\taub(t)\), \(\Delta_\taub(t)\), and \(K_\taub(t)\) remain independent of the reservoir temperatures.

These findings underscore the importance of considering finite-time effects in real-world thermal processes. Future research could explore more complex quenching protocols, including temperature-dependent heat transfer coefficients and nonlinear memory effects, to gain a deeper understanding of relaxation dynamics in nonequilibrium systems.

\acknowledgments
This work was inspired by discussions during my participation in the long-term workshop ``Frontiers in Non-equilibrium Physics 2024'' (YITP-T-24-01). I sincerely appreciate the warm hospitality extended to me by Prof.~Hayakawa during my stay at the Yukawa Institute for Theoretical Physics, Kyoto University, in July 2024.
I acknowledge financial support from Grant No.~PID2020-112936GB-I00 funded by MCIN/AEI/10.13039/501100011033, and from Grant No.~GR24022 funded by Junta de Extremadura (Spain) and by
European Regional Development Fund (ERDF) ``A way of making Europe''.

\section*{Data availability}

The data that support the findings of this article are not
publicly available because this publication is theoretical work.
The data are available from the author upon reasonable
request.

\appendix
\section{Limit of no time delay ($\tau=0$)}
\label{appA}
In the case $\tau=0$, Eqs.~\eqref{5} become
\begin{subequations}
\beq
\label{A5a}
\lim_{\tau\to 0}\wEE_0(s)=\frac{1}{s+1},
\eeq
\beq
\label{A5b}
\lim_{\tau\to 0}\wFF_\taub(s)=\frac{1}{\left(s+1\right)\left(s+\taub^{-1}\right)}.
\eeq
\end{subequations}
In real time,
\begin{subequations}
\beq
\lim_{\tau\to 0}\EE_0(t)=e^{-t},
\eeq
\beq
\lim_{\tau\to 0}\FF_\taub(t)=\frac{\taub}{1-\taub}\left(e^{-t}-e^{-t/\taub}\right).
\eeq
\end{subequations}
Therefore,
\beq
\label{11a}
\lim_{\tau\to 0}\EE_\taub(t)=\frac{e^{-t}-\taub e^{-t/\taub}}{1-\taub}.
\eeq
\section{Regime of quasi-instantaneous quenches ($\taub\ll 1$)}
\label{appB}
Let us first rewrite Eq.~\eqref{5b} as
\beq
\wFF_\taub(s)=\frac{\taub}{1-\taub e^{-s\tau}}\left(\frac{s^{-1}}{1+s^{-1}e^{-s\tau}}-\frac{1}{s+\taub^{-1}}\right).
\eeq
If $\taub\ll 1$, we can neglect $\taub e^{-s\tau}$ against $1$, i.e.,
\bal
\wFF_\taub(s)\approx&\taub\left(\frac{s^{-1}}{1+s^{-1}e^{-s\tau}}-\frac{1}{s+\taub^{-1}}\right)\nn
=&-\taub\left[\sum_{n=0}^\infty  (-s)^{-(n+1)}e^{-ns\tau}+\frac{1}{s+\taub^{-1}}\right].
\eal
The inverse Laplace transform is
\beq
\FF_\taub(t)\approx \taub\left[\sum_{n=0}^{\lfloor t/\tau\rfloor}\frac{(n\tau-t)^n}{n!}-e^{-t/\taub}\right].
\eeq

\bibliography{C:/AA_D/Dropbox/Mis_Dropcumentos/bib_files/Granular}

\end{document}